# Reliability comparison of AlGaN/GaN HEMTs with different carbon doping concentration

Z. Gao*, M. Meneghini, F. Rampazzo, M. Rzin, C. De Santi, G. Meneghesso, E. Zanoni

*Dipartimento di Ingegneria dell'Informazione, Università di Padova, Via Gradenigo 6/A, 35131 Padova*

Abstract

The reliability of AlGaN/GaN HEMTs adopting Fe and C co-doping, with high and low carbon doping concentration was investigated by means of different stress tests. Firstly, DC and pulsed I-V characterization at room temperature are discussed, then drain step stress tests at different gate voltages are compared, afterwards, the constant stress at different bias points are discussed. Results show that the high C HEMTs showed reduced DIBL, smaller leakage current, as well as decreased electric field, leading to an improved robustness during on-state stress testing, with respect to the reference ones. Failure modes during constant voltage stress consists in a decrease of drain current and transconductance, accelerated by temperature and electric field.

## 1. Introduction

GaN-based High Electron Mobility Transistors (HEMTs) have drawn great attention due to their potential for high temperature, high power and high frequency applications to radar amplifiers or modern telecommunication systems as 5G [1]–[4]. However, in scaled gate-length devices, leakage current, short channel effects and current collapse are still issues which may affect device performance [5] and reliability [6]. In order to reduce short-channel effects and off-state leakage currents, compensation doping, such as carbon [7] or iron [8] is introduced. Many studies on either iron or carbon doping designs have been published [9], [10]. Fe acceptors have been reported to be related to the presence of traps located at 0.72 eV or 0.63 eV below conductance band [11], [12], but it may lead to kink effects in the devices under certain conditions [13]. Carbon doping is used both for power switching and RF GaN HEMTs and it has been proved to be effective in reducing leakage current and increasing breakdown voltage [14]. Due to its self-compensation effect [15], [16], it can replace Ga or N, working as donor or acceptor trapping centres.

In this work we present data on the on-wafer reliability of 0.15 μm GaN HEMTs for microwave applications adopting Fe and C co-doping, with the same Fe doping profile, but different levels of carbon doping within the GaN buffer. The experimental and the electrical tests are described in Section 2. The results of the electrical characterization of the devices during each stress test are shown in Section 3, leaving the discussion and comparison at the end of that section. Finally, the main conclusions of the work are summarized in section 4.

## 2. Experimental details

Tested devices were fabricated on AlGaN/GaN heterostructures grown on SiC wafer, with gate length of 0.15 μm and gate width of 100 μm. The two different epi substrates, hereafter identified as "reference" and "High C", respectively, for the low C doping and high C doping profile, were processed within the same batch using a standard RF GaN HEMT process. Both wafers have the same iron doping profile, with maximum Fe concentration close to $2\times10^{18}$ $cm^{-3}$, however, the carbon doping concentration is close to $2\times10^{16}$ $cm^{-3}$ for the "reference" wafer and $8\times10^{16}$ $cm^{-3}$ for the "High C" wafer.

Standard DC, pulsed IV and electroluminescence (EL) spatial mapping tests were conducted on untreated devices at the beginning. Then a series of electrical stress tests was carried out. In order to identify the bias limits of the devices, drain step stress tests at three bias points (in off-state, semi-on state and on-state conditions) were carried out on at least two devices. At each bias point, drain voltage was increased from 0 V to devices catastrophic breakdown in 5 V step, two minutes long each. During each stress step, drain and gate currents as well as EL intensity were monitored. After each stress step, the device was kept unbiased for 5 minutes, a standard DC characterization was taken afterwards. Once viable bias points were identified from the drain step stress results, specific bias points for 24 hours DC tests were defined, and one device was tested at each bias points.

* Corresponding author. gaozhan@dei.unipd.it
Tel: +39 049 827 7625

## 3. Results and discussion

### *3.1. Preliminary characterization*

#### *3.1.1 DC characterizations*

DC output and subthreshold characteristics of the Reference and High C devices are shown in Fig.1. High C devices show reduces Drain Induced Barrier Lowering (DIBL) effects (40 mV/V for High C and 81 mV/V for reference, respectively), possibly due to the better charge confinement achieved thanks to high carbon doping concentration [17].

Gate diode ideality factor, calculated from the forward current measurements of the gate diode, is close to 2.1, for both reference and High C devices.

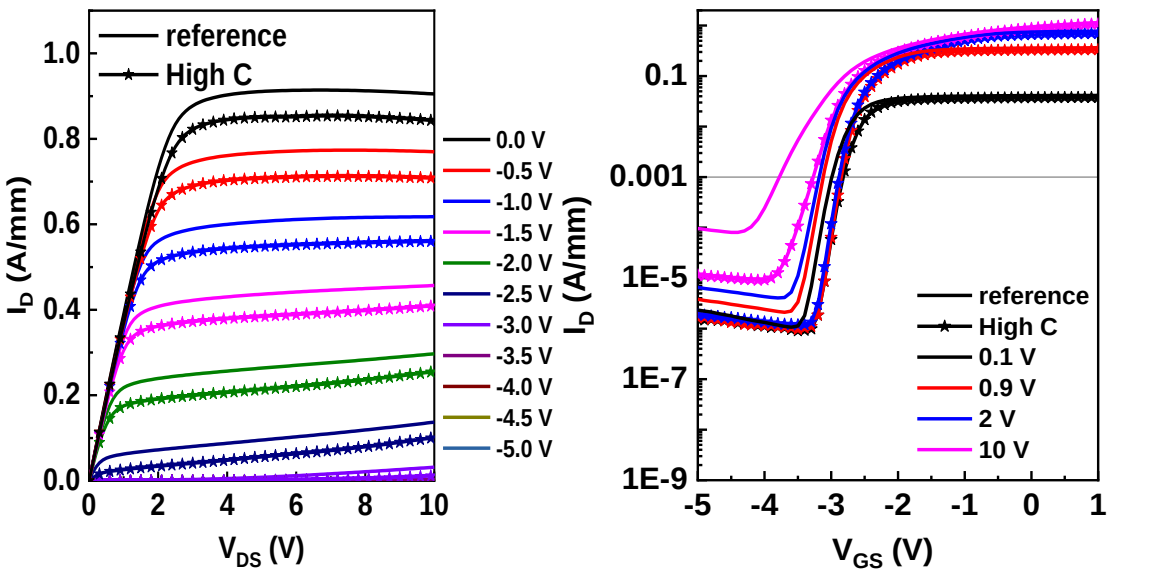

Fig.1. DC (a) output $I_D$-$V_D$ and (b) transfer characteristics comparison between reference and High C devices

EL images were taken at various drain and gate voltages, the EL intensity as a function of $V_D$ and $V_G$ are shown in Fig. 2. The EL intensity follows a bell shape, similar to that described in [18]. The High C devices showed reduced EL intensity in the semi-on and on state ($V_G$ > -1.5 V) when compared with the reference device, in accordance with smaller $I_{DSmax}$ during DC characterization. High C devices also showed a lower EL/$I_D$ ratio in on-state (Fig. 2b), a result consistent with a possible reduction of electric field in the channel at increasing $V_{GS}$ (other effects which may contribute to decrease this ratio include increased interface scattering and worsening of carrier transport properties due to device self-heating).

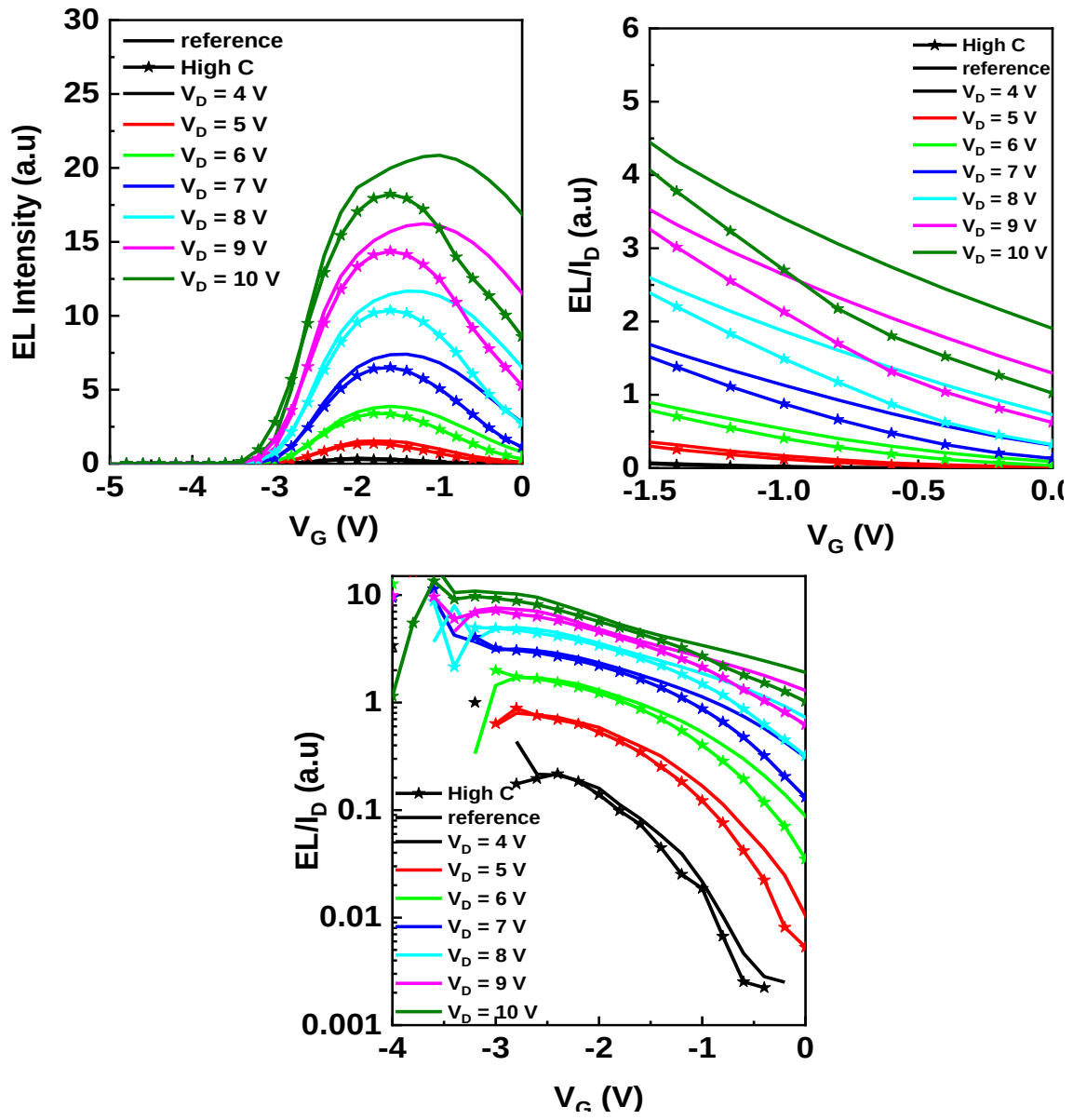

Fig. 2. (a) EL intensity and EL/$I_D$ (b) from -1.5 V to 0 V (c) from -4 V to 0 V as a function of gate voltage and drain voltage for reference and High C devices.

Pulsed IV characterization were carried out on both devices, at quiescent bias points $V_{GS,Q}$, $V_{DS,Q}$ = (0 V, 0 V) and $V_{GS,Q}$ = -6 V and $V_{DS,Q}$ = 0 V, 15 V and 25 V, with a pulse width of 1 μs and a pulse period of 100 μs; the related IV curves are shown in Fig. 3. Results showed that the two kinds of devices showed similar drain lag at $V_{DS,Q}$ = 15 V. But the drain lag at $V_{DS,Q}$ = 25 V of the reference HEMT is larger than the High C HEMT, possibly due to enhanced electron injection in the buffer. However, the GaN: High C device showed a larger gate lag and larger $V_{th}$ positive shift (+0.2 V) compared with the reference device, which could be due to charge trapping process under the gate, located in the buffer [19]. The slightly larger gate lag of the High C device may be correlated to a higher density of traps in the buffer, possibly related to C doping, since barrier and surface are the same for the two analysed wafers.

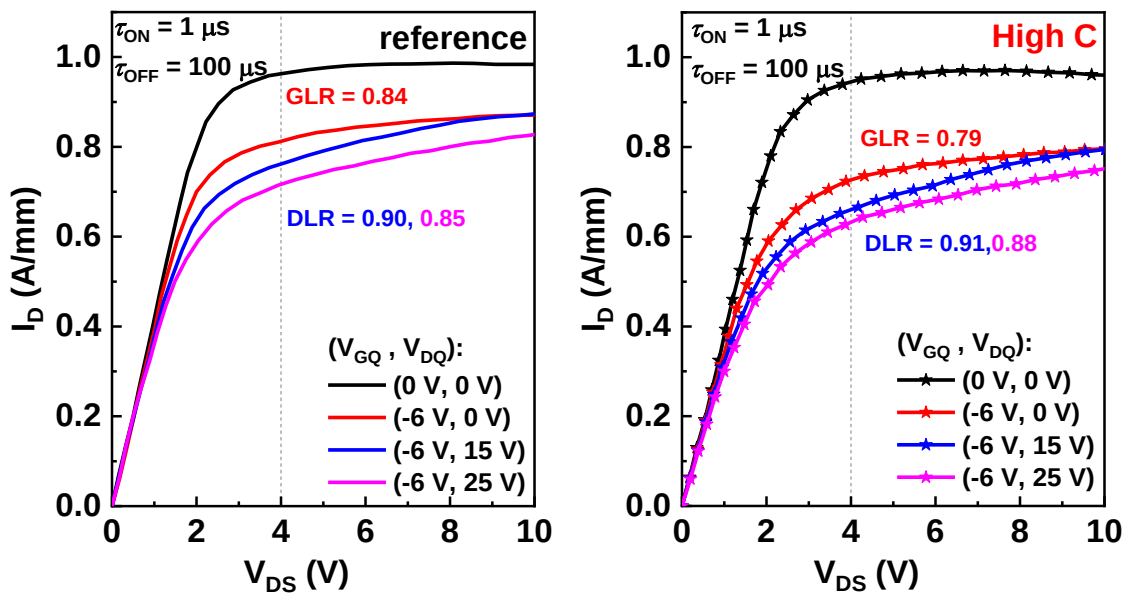

Fig. 3. Pulsed IV characteristics for (a) reference and (b)

High C devices at quiescent bias points (0 V, 0 V), (-6 V, 0 V), (-6 V, 15 V) and (-6 V, 25 V).

### *3.2. Drain Step Stress*

Drain step stresses were carried out at three bias conditions, off-state ($V_G$ = -7 V), semi-on states, where the EL bell peaks lies ($V_G$ = -2 V), and on-state ($V_G$ = 0 V), with at least two devices from each wafer at each bias condition, in order to compare failure modes and mechanisms of the two kinds of devices.

#### 3.2.1 off-state step stress

Both wafers were biased with gate voltage of -7 V, at off-state, and the drain bias was increased from 0 V to breakdown, with a step of 5 V. Both devices showed catastrophic breakdown at 60 V. At the stress step before failure, $V_{th}$ shifted positively in the Reference devices, by about 0.15 V; however, it shifted negatively in the High C device, by -0.2 V. The on resistance ($R_{ON}$) increase in the High C device is 15% smaller than in the reference devices. Both devices showed close to 10% decrease of maximum transconductance ($g_{m,max}$), as shown in Fig. 4. However, there is no significant increase in gate/drain leakage current in both devices. A critical failure voltage ($V_{critical}$) was defined as the minimum $V_{DS}$ inducing a 10% decrease/increase in any of the main DC parameters, i.e. $I_{DS,max}$, $g_{m,max}$ and $R_{ON}$. $V_{critical}$ of the High C device is about 60 V (the same to the breakdown voltage), and that of the reference device is 45 V (referred to $R_{ON}$ increase); significantly. A negative $V_{th}$ shift (-0.2 V), together with $I_{DSmax}$ increase, was observed in the High C devices, while $V_{th}$ moved positively (0.15 V) in reference devices, see Fig. 4.

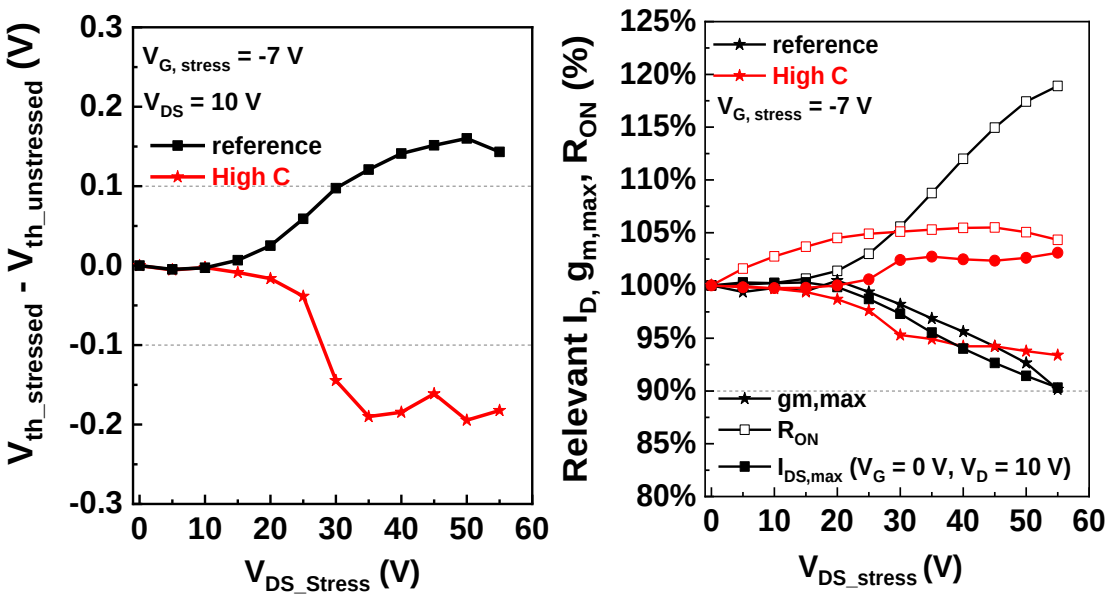

Fig. 4 (a) normalized $V_{th}$ and (b) $I_{DS,max}$, $R_{ON}$ amd $g_{m,max}$ during off-state stress at $V_G$ = -7 V of reference and High C devices.

#### 3.2.2 semi on-state step stress

We compared relative changes of the DC characteristics by applying a "semi-on" state step stress at a bias point $V_G$ = -2 V, which is close to the EL peak shown in Fig. 2, i.e. to maximum hot electron effects. Both devices showed catastrophic failure at 50 V, and showed firstly negative, then positive $V_{th}$ shift. Comparing the $I_{DS,max}$, $g_{m,max}$ and $R_{ON}$ degradation, the failure voltage is about 35 V and 40 V (referred to $I_{DS,max}$ decrease) and for the High C and reference device, respectively, as shown in Fig. 5.

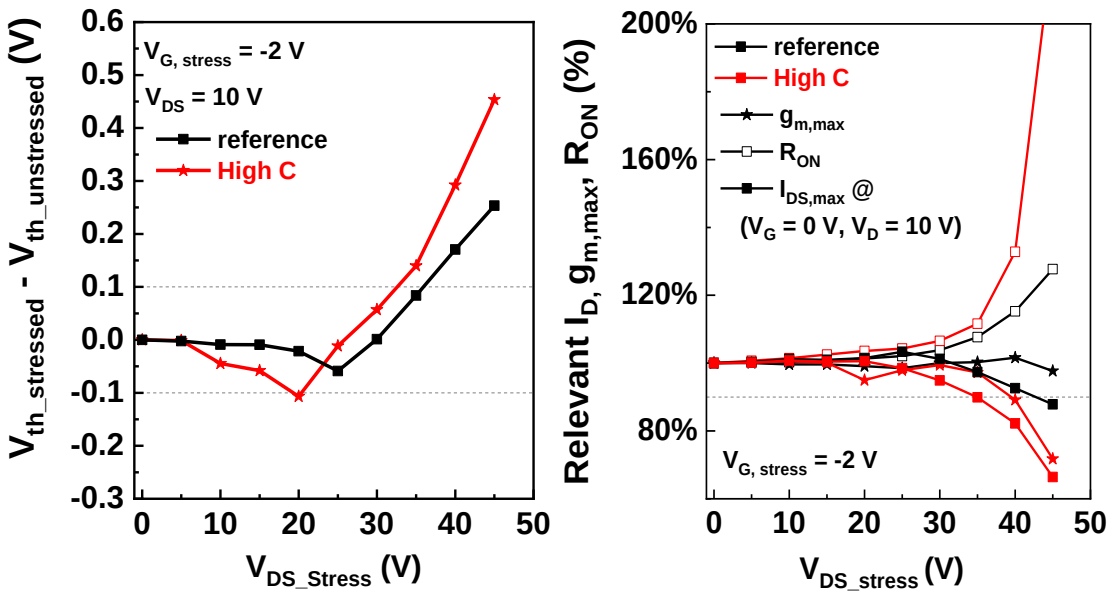

Fig. 5 (a) normalized $V_{th}$ and (b) $I_{DS,max}$, $R_{ON}$ amd $g_{m,max}$ during semi on-state stress at $V_G$ = -2 V of reference and High C devices.

#### 3.2.3 on-state step stress

With regard to the stress at on-state, at bias point $V_G$ = 0 V, both devices failed catastrophically at 45 V. However, for what concerns $I_{DS,max}$ and $g_{m,max}$, the High C devices failed at 35 V, while reference devices presented a critical failure voltage of about 20 ~ 25 V, as shown in Fig. 6.

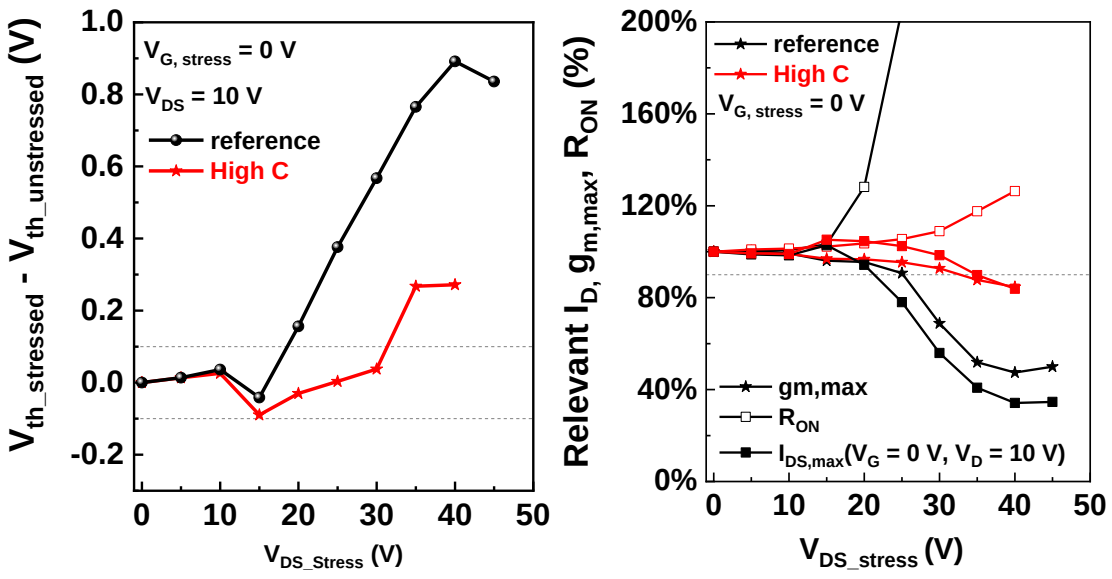

Fig. 6. (a) normalized $V_{th}$ and (b) $I_{DS,max}$, $R_{ON}$ and $g_{m,max}$ during on-state stress at $V_G$ = 0 V of reference and High C devices.

Similar to the stress at $V_G$ = -2 V, both devices showed firstly $V_{th}$ negative shift, followed by a remarkable positive shift. The amount of positive $V_{th}$ shift, however is much larger in the reference devices with respect to the High C ones (approximately 900 mV and 300 mV, respectively).

Failure voltage ($V_{critical}$) and breakdown voltage ($V_{BR}$) are identified on the $I_D$ vs $V_{DS}$ curves in Fig. **7**. The two epitaxial structures show similar breakdown voltage levels (unaffected by C-doping level), but their behavior is different concerning parametric degradation during step-stress tests: High C devices show better critical voltages both in off-state and on-state, whereas comparable levels of degradation are found in semi-on state, where hot-electron effects are close to their maximum ($V_{GS}$ = -2 V, in Fig. 2).

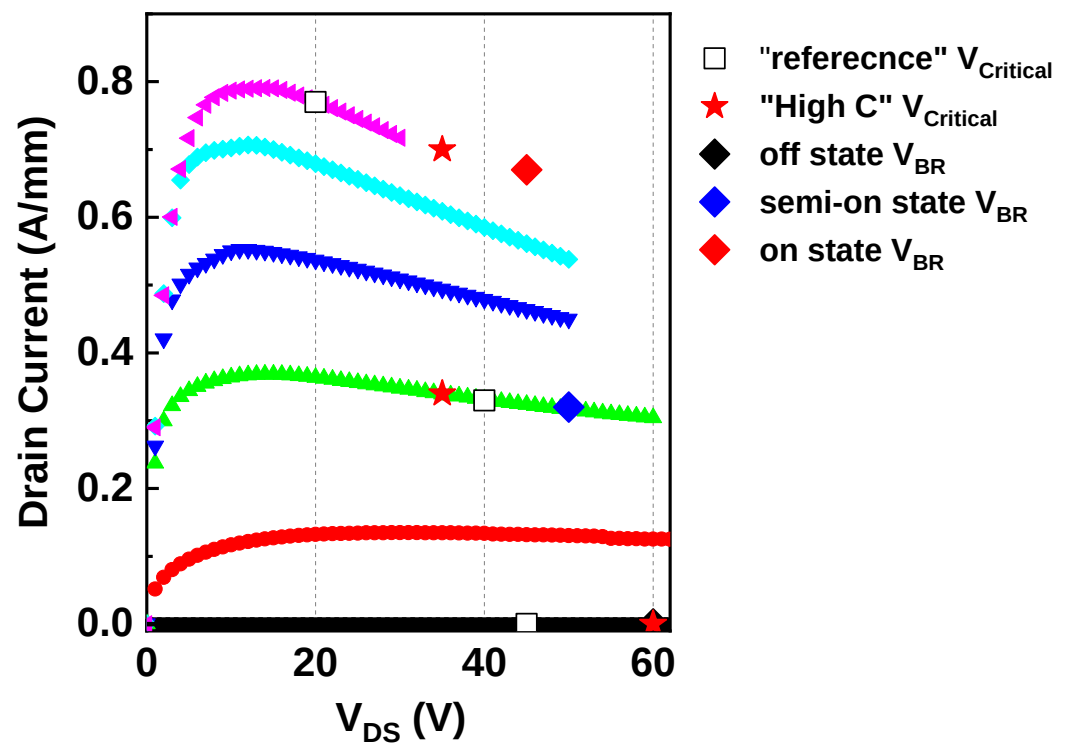

Fig. 7. $V_{BR}$ and $V_{critical}$ comparison between High C and reference devices.

*3.3. 24-hours DC Stress*

Based on the results achieved from step stress tests, we have carrier out a series of 24-hours stress tests, to evaluate the devices reliability on a short time scale. A series of bias points was chosen at off-state, with $V_G$ = -3 V and $V_D$ from 10 V to 25 V with 5 V step, a new fresh device was used for each stress condition. DC characteristics degradation rate was summarized after the 24 hours' stress. $\Delta I_{DSmax}$ and $\Delta g_{mmax}$ as a function of $V_{DS}$ were summarized, as shown in Fig. 8. No significant difference was found for the two technologies.

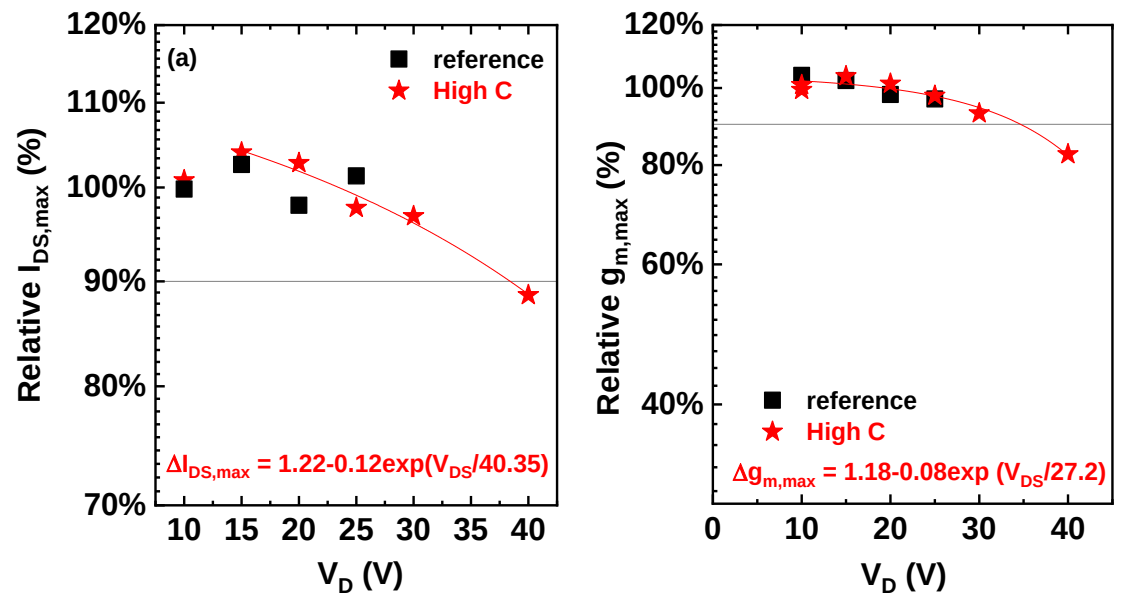

Fig. 8. (a) $I_{DSmax}$ and (b) $g_{m,max}$ during off-state constant voltage stress at $V_{GS}$ = -3 V and several $V_{DS}$.

In order to further compare the degradation effects of the two kinds of devices at stress conditions with relatively high dissipation power ($P_{Diss}$) /channel temperature ($T_{ch}$), several bias points were chosen in the on-state, with -1 V and 0 V, and $V_D$ from 10 V to 25 V with a step of 5 V. Thermal resistance values estimated, using the technique described in [20], are 21.3 K*mm/W and 19.0 K*mm/W, for the reference and High C devices, respectively.

Fig. 9 shows the dependence of drain current and transconductance for the devices. DC tested for 24 hours, each test either at $V_{GS}$ = -1 V or 0 V, and at a $V_{DS}$ value ranging from 10 V to 25 V with a 5 V step. In Fig. 9, $\Delta I_{DSmax}$ and $\Delta g_m$ are plotted as a function of the DC dissipation power $P_{Diss}$ and channel temperature $T_{ch}$.

In on state conditions, two main cause of failure could be considered, electric field and channel temperature. During stress, the reference device showed significantly higher decrease of $I_{D,max}$ for all temperature values, while $g_{m,max}$ is more comparable, as shown in Fig. 9.

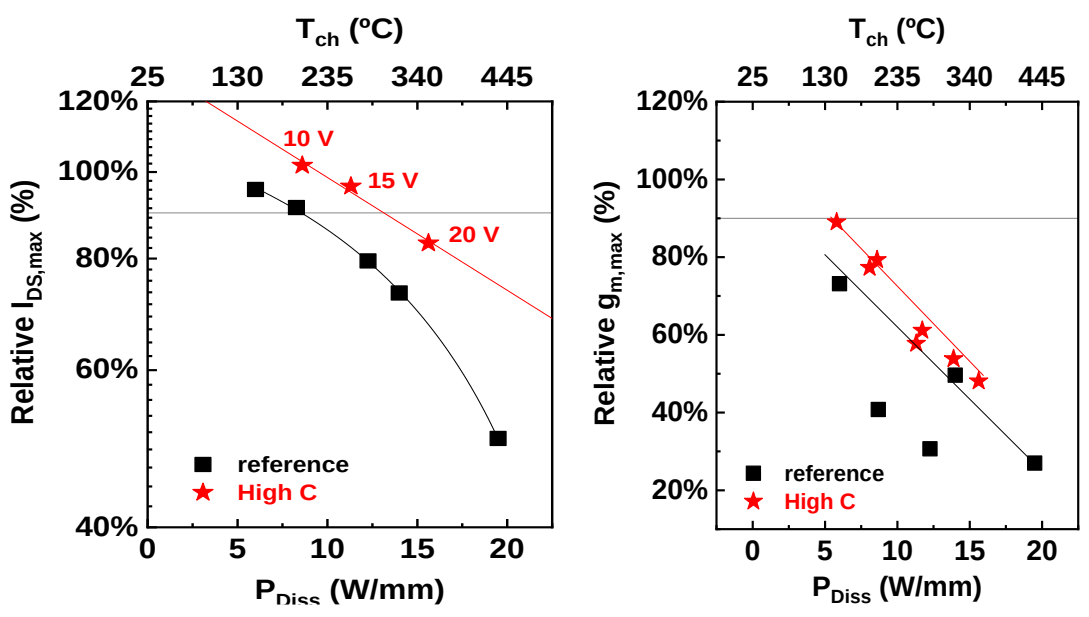

Fig. 9 (a) $I_{DSmax}$ and (b) $g_{m,max}$ decrease during on-state constant voltage stress varying $V_{D,Stress}$.

## 4. Conclusions

On-wafer robustness and short-term reliability of 0.15 µm Fe and C co-doped AlGaN/GaN HEMTs, with standard and high carbon doping concentration in the buffer, has been compared. Results can be summarized and tentatively explained as follows:

(a) in untreated devices, high C doping is effective in reducing short channel effects, like low subthreshold slope, DIBL, and source-drain leakage current (Fig. 1), at a cost of a slightly higher current collapse., which could possibly be due to the enhanced trapping related to C-related deep levels, as shown in Fig. 3.

(b) Electroluminescence measurements suggested a decrease of the electric field and hot-electron effects (measured by the EL/$I_D$ ratio) in the high C devices for $V_{GS}$ > -2 V (on-state), while hot electron effects are comparable in off-state and semi-on state (Fig. 2).

(c) The dominant failure mode consists in a positive threshold voltage shift, leading significant drain current decrease, possibly due to hot-electron trapping in the buffer at the gate edge towards the drain, which also affects peak transconductance. In on-state, as demonstrated by EL/$I_D$ measurements, hot electron effects are stronger in reference devices, and this explains the larger degradation observed at $V_{GS}$ = -1 V or 0 V, in Fig. 6 and Fig. 9. For all channel temperatures in on-state, degradation of reference device is always larger when compared to the High C ones, demonstrating that the difference is actually related to the different electric field within the epi layer. No difference is found for the two technologies concerning degradation in semi-on state, as hot-electron effects are comparable in this bias condition, Fig. 2.

(d) Finally, the different behaviour of reference and High C devices observed during off-state step stress can be explained by the injection of holes taking place in pinch off conditions at high drain bias, which is related to the modulation of charge at $C_N$ acceptors, possibly related leading to negative $V_{th}$ shifts, as shown in [13]. This might even have a beneficial effect on device reliability, by contracting negative electron charge trapping.

## Acknowledgements

Support by EUGANIC project under the EDA Contract B 1447 IAP1 GP, by the EC Horizon 2020 ECSEL project 5G_GaN_2, and by the ESA ESTEC project RELGAN is gratefully acknowledged.